# The properties of the nitrogen-vacancy center in milled chemical vapor deposition nanodiamonds

Alessandro Mameli[1], Giannis Thalassinos[1], Marco Capelli[1], Johannes Ackermann[2], Edwin Mayes[1], Hiroshi Abe[3], Takeshi Ohshima[3,4], Tingpeng Luo[5], Volker Cimalla[5], Peter Knittel[5], Brant Gibson[1], Jan Jeske[5], Nikolai Dontschuk[6], Anke Krueger[2,7], Alastair Stacey,[1,8] Alexander Healey[1*], Philipp Reineck[1*]

[1] *School of Science, RMIT University, Melbourne, VIC 3001, Australia*
[2] *Institute of Organic Chemistry, University of Stuttgart, Pfaffenwaldring 55, 70569, Stuttgart, Germany*
[3] *National Institutes for Quantum and Radiological Science and Technology, Takasaki, Gunma, 370-1292, Japan*
[4] *Department of Materials Science, Tohoku University, Aoba, Sendai, Miyagi 980-8579, Japan*
[5] *Fraunhofer Institute for Applied Solid State Physics IAF, Tullastraße 72, 79108 Freiburg im Breisgau, Germany*
[6] *School of Physics, University of Melbourne, Parkville, Victoria 3010, Australia*
[7] *Center for Integrated Quantum Science and Technology, 70569, Stuttgart, Germany*
[8] *Princeton Plasma Physics Laboratory, Princeton University, Princeton, NJ 08540, USA*

\* *alexander.healey2@rmit.edu.au, philipp.reineck@rmit.edu.au*

**Abstract**
Fluorescent nanodiamonds (FNDs) containing negatively charged nitrogen-vacancy (NV$^-$) centers are vital for many emerging quantum sensing applications from magnetometry to intracellular sensing in biology. However, developing a scalable fabrication method for FNDs hosting color centers with consistent bulk-like photoluminescence (PL) and spin coherence properties remains a highly desired but unrealized goal. Here, we investigate optimized ball milling of single-crystal diamonds produced via chemical vapor deposition (CVD) and containing 2 ppm of substitutional nitrogen and 0.3 ppm of NV$^-$ to achieve this goal. The NV charge state, PL lifetime, and spin properties of bulk CVD diamond samples are directly compared to milled CVD FNDs and commercial high-pressure high-temperature (HPHT) FNDs. We find that on average, the relative contribution of the NV$^-$ charge state to the total NV PL is lower and the NV PL lifetime is longer in CVD FNDs compared to HPHT FNDs, both likely due to the lower $N_s^0$ concentration in CVD FNDs. The CVD bulk and CVD FNDs on average show similar average $T_1$ spin relaxation times of 3.2 ± 0.7 ms and 4.7 ±1.6 ms, respectively, compared to 0.17 ± 0.01 ms for commercial HPHT FNDs. Our results demonstrate that ball milling of CVD diamonds enables the large-scale fabrication of NV ensembles in FNDs with bulk-like $T_1$ spin relaxation properties.

**Introduction**
Color centers in diamond are at the forefront of emerging quantum technologies, from quantum networks [1] and magnetometry [2, 3] to nanoscale sensing in biology [4]. While many technologies are based on bulk diamond samples, fluorescent nanodiamonds (FNDs) offer some unique advantages. FNDs can be integrated into other materials [5-8] and devices [9], and delivered to and accessed in many nanoscale environments within biological and chemical systems [4]. FNDs thus enable nanoscale quantum sensing and imaging in locations and for platforms that are inaccessible for bulk diamonds. However, the reliable creation of color centers in nanodiamonds with bulk-like spin properties at scale remains a challenge and limits the sensitivity and consistency of FND-based quantum sensing.

While many promising color centers are explored [10], the majority of emerging FND-based quantum sensing technologies employ the NV$^-$ center as the nanoscale sensor. $T_1$ spin relaxometry and optically detected magnetic resonance (ODMR) measurements are the most common sensing approaches for magnetic noise detection (e.g. radicals in biology and chemistry), and magnetometry and thermometry, respectively. For $T_1$ relaxometry, sensitivity scales with the relaxation rate $1/T_1$, and for ODMR-based sensing, sensitivity is proportional to $\frac{\Delta\nu}{C\sqrt{R}}$, where $\Delta\nu$ is the ODMR linewidth, $C$ the ODMR contrast, and $R$ the average PL photon detection rate. ODMR linewidth is proportional to the transverse spin relaxation rate $1/T_2^*$. Hence, the goal is to maximize the $T_1$ and $T_2$ times, ODMR contrast, and PL brightness simultaneously.

The most common approach for the fabrication of FNDs for quantum sensing applications remains the milling of irradiated and annealed high-pressure high-temperature (HPHT) microparticles [11]. These processes have been optimized over the past decade [12] but some fundamental challenges associated with HPHT diamonds remain. The typically high substitutional nitrogen ($N_s^0$) content of ~ 100 ppm in HPHT diamonds results in FNDs with high PL brightness and stabilizes the NV$^-$ charge state, but limits the transverse spin relaxation time ($T_2$) due to NV dephasing caused by the nitrogen spin bath [13], thereby limiting sensitivity. Typical $T_1$ spin relaxation times in HPHT FNDs are on the order of 100 µs [14], compared to 2-5 ms in bulk samples with $N_s^0$ concentrations < 100 ppm, where $T_1$ is dominated by phonon-induced relaxation at room temperature [15]. The large variation of $N_s$ concentrations in different growth sectors of HPHT diamonds likely contributes to the great variability in single FND properties [16]. Finally, the milling process itself likely creates additional defects and strain in the crystal, irrespective of the starting material [12], and can create surface-related magnetic noise, for example, from residual metal impurities from the milling balls [17] or the increase of surface defects [18]. However, due to the wide variety of potential noise sources in typical milled FNDs, it is difficult to isolate the contribution of the milling process.

Hence, alternative approaches to create FNDs tailored for quantum sensing applications are highly sought after. One approach is to use chemical vapor deposition (CVD) to grow individual FNDs on a solid substrate [19] or to etch particles from CVD thin films [20]. CVD enables good control over the concentration and homogeneity of dopants like nitrogen in the crystal and can yield single particles with bulk-like spin coherence properties [19, 20]. However, both processes are difficult to scale to the gram-scale of material required for most real-world applications, i.e. don't take advantage of the relative scalability of CVD diamond growth. Hence, combining the controlled engineering of defects offered by CVD with the scalability of optimized milling processes is a possible route for the creation of FNDs with improved spin properties. Recently, Ohshimi et al reported improved NV coherence properties in diamond particles milled from $^{12}$C-enriched CVD diamonds [21]. FNDs milled from single-crystal CVD diamond containing ~ 0.1 ppm $N_s^0$ with $T_2$ coherence times of over 400 µs (using dynamical decoupling) were reported for individual particles by Wood et al. [22]. The particles had a mean size of less than 100 nm and contained single or a few NV centers per particle. While the single-photon emission properties of single or few NVs may offer unique advantages for certain applications, many applications require FNDs with higher NV concentrations per FND and, consequently, higher PL brightness to enable PL detection in the presence of background fluorescence, for example, in biological samples. Furthermore, $N_s^0$ concentrations on the order of 0.1 ppm mean that only around 10 electron donors are present on average in a 100 nm FND, orders of magnitude below the anticipated number of surface acceptor states. This disparity suggests a large fraction of NVs will not be in the NV$^-$ charge state. Hence, it is important to explore intermediate nitrogen concentrations in FNDs that balance PL brightness and the NV charge state without the detrimental effects of $N_s^0$ concentrations above 100 ppm.

Here, we investigate the NV PL and spin properties of FNDs milled from single-crystal CVD diamonds containing ~2 ppm $N_s^0$ and ~0.3 ppm NV$^-$ centers. We directly compare the CVD bulk sample with milled CVD FNDs around 100 nm in size and commercially available HPHT FNDs of similar size. We investigate the NV charge state ratio, PL lifetime, ODMR properties and $T_1$ spin relaxation times in all samples. The properties of single particles and average properties of particle ensembles are analyzed. We find a lower average contribution of NV$^-$ to NV$^0$ ratio and a longer average PL lifetime in the milled CVD FNDs relative to the CVD bulk and the HPHT FNDs. The ODMR zero-field splitting parameter $D$ is reduced by about 2.5 MHz for both FND samples compared to the bulk, likely due to milling-induced strain. The ODMR zero-field splitting parameter $2E$, related to electric field and strain effects, is increased in both FND samples relative to the bulk sample. Here, the increase for the CVD FND sample is lower than that for the HPHT FNDs, likely due to a lower concentration of charged defects in the CVD FNDs. Lastly, we find that the CVD bulk and CVD FND samples show a similar average $T_1$ spin relaxation time of 3.2 ± 0.7 ms and 4.7 ±1.6 ms, respectively, compared to 0.17 ± 0.01 ms for the HPHT FNDs. Our results show that bulk-like $T_1$ relaxation times can be preserved in FNDs with intermediate NV concentrations and highlight the need for improved defect engineering to control the NV charge state in CVD FNDs.

**Experimental**
**Material synthesis and processing**
*CVD synthesis*. Three samples were grown on 3 × 3 mm² HPHT substrates (Element Six). The growth substrates were polished to a surface roughness of $R_a$ < 0.5 nm and a thickness of approximately 200 µm. Prior to growth, all substrates underwent an extensive cleaning process, which included boiling in a mix of nitric and sulfuric acid and several successive cleaning steps in organic solvents. Growth took place in a specially designed ellipsoidal CVD reactor with a 2.45 GHz, 6 kW microwave generator [23]. It was initiated by treatment in $H_2/O_2$ and $H_2$ plasma for 10 minutes each at 800°C. Growth took place at 210 mbar, between 800 °C and 830 °C, with 2.67% methane in hydrogen and a small amount of nitrogen gas, leading to a nitrogen/carbon ratio of 6500 ppm in the plasma. After 45 hours of growth, a thickness of approximately 500 µm was achieved. Finally, the growth side was polished and cleaned in the same way as before the growth, and the CVD diamond was removed from the HPHT substrate.
*NV creation*. Bulk CVD diamond samples were irradiated with 2 MeV electrons to a fluence of 1 × 10$^{18}$ e$^-$ cm$^{-2}$ and annealed in vacuum (900 °C, 2 h).
*Sample milling*. Two of the three bulk CVD samples were placed in a stainless-steel milling cup containing 15 stainless-steel milling balls (4 mm diameter). Subsequently, 5 mL of isopropanol was added, and the milling cup was sealed and mounted in a vibrating ball mill (Fritsch Pulverisette 23). The diamond was milled for 6 hours at a vibration frequency of 50 Hz. After milling, the contents of the milling cup were transferred into a 500 mL round-bottom flask using additional isopropanol to rinse the cup. The steel milling balls were removed using a magnet, and the isopropanol was evaporated under reduced pressure. The flask was then placed in a drying cabinet at 120°C for 1 hour to ensure complete removal of residual isopropanol. To the dried diamond material, 100 mL of concentrated sulfuric acid (96%) was added slowly under stirring to minimize foaming caused by the reaction between the acid and residual metal debris. Once gas evolution had ceased, 50 mL of nitric acid (65%) was added. The mixture was stirred continuously at 140°C for 12 hours. Afterward, heating and stirring were stopped, allowing the diamond to settle at the bottom as the mixture cooled to room temperature. Once cooled, most of the acid mixture was removed by decantation. The diamond material was washed through repeated centrifugation at 15,000 rpm for 1 hour, followed by decantation of the supernatant and replacement with distilled water. The diamond material was dispersed by sonication between centrifugation cycles. This washing process was repeated until the supernatant reached a neutral pH.

**Materials characterization**
*UV-Visible absorption spectroscopy*: The sample transmittance was measured at room temperature from 200–800 nm using a UV/Vis/NIR spectrometer (PerkinElmer Lambda 950). The absorption coefficient was calculated from the measured transmittance and sample thickness using the Beer–Lambert law. The spectra can be analysed to understand dopant concentrations, particularly the single substitutional $N_s^0$, also known as P1 centres, can be quantified via their characteristic absorption at 270 nm and the corresponding absorption cross section [24]. $NV^-$ and $NV^0$ absorption spectra can also be seen for high enough concentrations via UV-Vis spectroscopy.
*Fourier-transform infrared (FTIR) spectroscopy:* FTIR spectra of the CVD bulk samples were acquired using a Invenio-R spectrometer (Bruker, USA).
*Birefringence measurements*: The map of birefringence ($\Delta n$) for the sample was obtained from spatially resolved measurements of the optical retardation ($\Gamma$) using a polarimeter (Ilis StrainMatic M4/90.50 Zoom) as in [25]. The retardation corresponds to the phase difference between the two optical axes with different refractive indices, and the birefringence was calculated as $\Delta n = \Gamma/d$.
*Photoluminescence (PL) measurements:* PL maps were measured in a home-built confocal microscope with 532nm pump laser power of 10 µW to confirm a homogeneous NV distribution in the sample. From the brightness, we determined the NV concentration. PL spectra were measured with a spectrometer (HORIBA iHR320) to determine the $NV^-$ and $NV^0$ charge state ratio. Details of the measurements are analogous to [26]
*Dynamic light scattering (DLS)*: DLS experiments were conducted using an ALV-5022F light scattering spectrometer with a laser wavelength of 633 nm (ALV-GmbH, Germany). FNDs were suspended in DI water at a concentration well below 0.1 mg mL$^{-1}$ in cylindrical glass cuvettes (inner diameter 8 mm) (LSI Instruments, Fribourg).
*Transmission electron microscopy (TEM)*: Samples were prepared for TEM via drop casting onto a Cu grid with a 10 nm formvar and 1 nm carbon membrane (GSFC200CU-SA, ProSciTech, Australia). Imaging was performed on a JEOL F200 TEM CFEG operating at 200 kV equipped with a Gatan Rio16 camera.

**Optical characterization and spin measurements**
*Sample Preparation*: Marked glass coverslip substrates were used as substrates for FND NP samples. The substrates were cleaned with acetone, isopropyl alcohol, and ultrapure water (Milli-Q, Merck), gently dried with $N_2$ flux, followed by a 10-minute cycle of ozone cleaning (UV Ozone Cleaner, Ossila). HPHT FND (NDNV100nmHi10ml, Adamas Nanotechnologies) were used as received. HPHT FNDs were spin-coated onto the glass substrate (2000 rpm) at a concentration of 0.1 mg mL$^{-1}$. CVD FNDs were drop cast due to their lower particle concentration in suspension.
*Fluorescence Imaging and Spectroscopy*: Fluorescence characterization was performed using a custom-built confocal fluorescence microscope equipped with a high-numerical-aperture objective lens (100×, NA = 0.9) and operated at room temperature. Excitation at 520 nm was provided by an 80 MHz pulsed laser (Fianium WhiteLase 400-SC, NKT Photonics). The emitted fluorescence was spectrally filtered using a 532 nm short-pass and a 532 nm long-pass filter, then collected through the same objective. The fluorescence signal was focused into a multimode optical fiber containing an internal beamsplitter that directed 75% of the signal to a spectrometer (SpectraPro with a PIXIS CCD camera, Princeton Instruments) and 25% to an avalanche photodiodes (SPCM-AQRH-14, Excelitas Technologies) detector. SI Figure S3 shows a detailed schematic of the confocal system.
Charge state analysis of NV centers was conducted following the dual-excitation protocol described by Thalassinos et al. [27], in which an $NV^0$ reference spectrum is acquired using a 405 nm laser excitation (Coherent Obis FP 1193825, Edmund Optics) and is subtracted from a spectrum showing both charge state contribution ($NV^0$ + $NV^-$) collected at 520 nm. Both signals were normalized for PL intensity <600 nm before subtraction to infer the $NV^-$ spectrum, then least-squares fitting was used to calculate the NV ratio. Filtering for 405 nm collection included a 405 nm laser line filter and a 450 nm long-pass filter to isolate the relevant emission.

Time-resolved fluorescence decay measurements were carried out using a photon counting board (TimeHarp 260, PicoQuant). A 532 nm dichroic mirror and long-pass filter (Semrock) were used to separate the excitation and emission signals. Image acquisition and sample scanning were conducted using a piezoelectric XYZ stage (PInanoXYZ, Physik Instrumente).

*Spin measurements:* Spin measurements were also performed using a confocal microscope. A 514 nm laser (Cobolt 06-MLD, Hubner Photonics) was used and gated using a MT110-A1.5-VIS acousto-optic modulator when required. Fluorescence was filtered (650 nm long pass) and collected onto the APD, and was recorded using an NI 9263 DAQ for ODMR measurements and time-resolved using a FAST ComTec MCS6a for $T_1$ measurements. Microwave (MW) radiation for spin state control was delivered through a custom printed circuit board. MW signals were generated by a Windfreak SynthNV Pro and amplified by a Minicircuits ZHL-16W-43+ amplifier. All spin measurements were sequenced using a PulseBlaster ESR Pro 500 MHz (SpinCore).

**Results and discussion**

Figure 1a) illustrates the synthesis and processing steps used to create the samples studied here. Three CVD diamond samples were grown as described above. Figure 1a), left, shows an image of an as-synthesized sample. NV centres were created via high-energy electron irradiation (2 MeV) to a fluence of $1 \times 10^{18}$ e$^-$ cm$^{-2}$ followed by annealing at 900 °C in vacuum for 2h. The samples were characterized using Fourier-transform infrared spectroscopy (FTIR), UV-Vis absorption spectroscopy, birefringence imaging, and PL spectroscopy to assess the sample quality and to ensure that the three samples have similar properties before milling. See SI Figure S1 and Table S1 for a direct comparison of the properties of the three CVD samples. All samples had similar concentrations of substitutional nitrogen ($N_s^0$), ranging from 1.81 to 2.05 ppm, and NV$^-$ concentrations between 0.25 and 0.28 ppm, determined via absorption spectroscopy [24]. The samples also show similar NV charge state ratios and NV $T_1$ spin relaxation times (see SI Figures S1c and d). Figure 1b) and 1c) show typical FTIR and UV-Vis absorption spectra, respectively, of the bulk diamond sample. The FTIR spectrum only shows absorption features associated with carbon-carbon bonds in the 1700 – 2600 cm spectral region. The UV-Vis absorption spectrum shows the characteristic absorption peak of $N_s^0$ ~267 nm and the NV$^-$ zero phonon line (ZPL) at 637 nm, accompanied by broader NV phonon sideband absorption between 480 and 600 nm. Together, these measurements confirm that $N_s^0$ and NV are the dominant defects in all three CVD samples and that they have consistent optical and NV PL properties.

Two of the three CVD samples were then milled into particles (CVD FND), and one bulk sample was kept to enable a direct comparison of bulk and nanoparticle samples. Bulk CVD diamond was milled using steel balls. Although these milling balls exhibit strong abrasion their use is favored over harder milling materials such as tungsten carbide or zirconia. The debris from steel balls can be efficiently removed using oxidizing mineral acids, which simultaneously removes surface-bound non-diamond carbon that was produced during the milling [28]. See the Experimental section for details.

Figure 1a), right, shows transmission electron microscopy (TEM) images of the milled and purified particles. The images show particles of a broad range of sizes, from several hundred nanometers to well below 100 nm in diameter. See SI Figure S2 for more TEM images. Figure 1d) shows volume-weighted and unweighted particle size distributions determined using dynamic light scattering (DLS). In agreement with TEM images, the unweighted size distribution shows the presence of a broad range of particle sizes from below 100 nm to above 1 μm, while the volume-weighted distribution reveals that most particles are ~100 nm in diameter. In the following, these particles will be referred to as CVD FND.

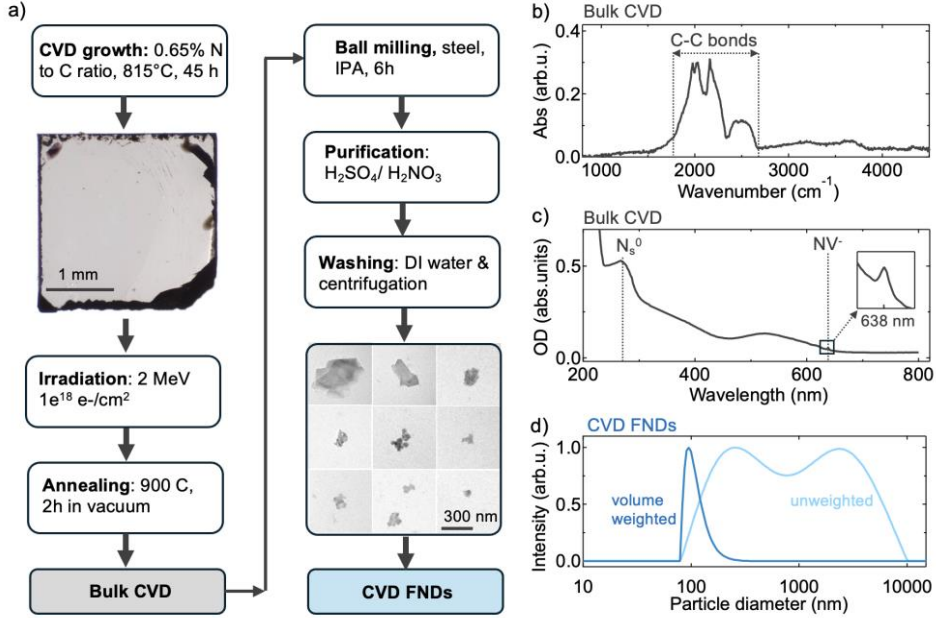

*Figure 1*. a) Flow chart of the sample synthesis and processing steps, and images of the as-synthesized bulk CVD samples (top left) and TEM images of the milled diamond particles (bottom right). Three bulk CVD samples were synthesized. Two of these were milled into particles (CVD FND), one was not milled to enable a direct comparison of bulk and nanoparticle samples. b) and c) Fourier-transform infrared (FTIR) spectrum (b) and UV-vis absorption spectrum of a bulk CVD sample after irradiation and annealing. d) Dynamic light scattering particle size distribution (unweighted intensity distribution) of the milled CVD FNDs suspended in DI water.

We first investigated the PL spectra and PL lifetime of the CVD bulk and CVD FND samples. We also included commercially available FNDs (Adamas Nanotechnologies, USA) made from HPHT diamond in our study as a reference point. This comparison is instructive since HPHT FNDs are the most common FNDs due to their high PL brightness and NV concentrations. It is important to note that the HPHT and CVD particles investigated here differ in several ways, summarized in Table 1. HPHT diamonds typically contain one to two orders of magnitude higher $N_s^0$ concentrations than the CVD samples investigated here. The HPHT FNDs were irradiated with a ~10× higher fluence (~$1 \times 10^{19}$ $e^-$ cm$^{-2}$) [29] compared to the CVD samples ($1 \times 10^{18}$ $e^-$ cm$^{-2}$). As a result of the higher nitrogen concentration and irradiation dose, the HPHT FNDs also have a higher concentration of NV centers of 1-3 ppm (according to the manufacturer) compared to the CVD FND sample (0.26 ppm).

| Sample | $N_s^0$ conc. (ppm) | Irradiation dose ($e^-$ / cm$^2$) | NV conc. (ppm) |
|---|---|---|---|
| Bulk CVD | 2.05 | $1 \times 10^{18}$ | 0.27 |
| CVD FND | 1.85 | $1 \times 10^{18}$ | 0.26 |
| HPHT FND | 100-200 | $1 \times 10^{19}$ | 1-3 |

*Table 1*. Summary of defect concentrations and irradiation doses for the investigated samples. $N_s^0$ and NV concentrations were determined through UV-Vis absorption spectroscopy for CVD samples and provided by the manufacturer for HPHT FND.

To study the PL properties of the particle samples, FNDs were spin-coated (HPHT FNDs) or drop-cast (CVD FNDs) onto a glass cover slip and imaged with a custom-built scanning confocal PL microscope using 520 nm ps-pulsed laser excitation and a 100× air objective. The CVD FNDs were drop-cast since the particle concentration in suspension was too low to yield a reasonable particle density on the substrate using spin coating. PL was collected through the same objective, separated from the excitation light with dichroic and long-pass filters, and fiber-coupled to acquire PL images, time-resolved PL decay traces, and PL spectra. We only considered diffraction-limited spots (around 300 nm in our setup) to avoid measuring large particles or aggregates. See Experimental section and SI Figure S3 for more details.

Figure 2a) shows average PL spectra for the studied samples normalized to the broad NV$^-$ phonon sideband peak at ~690 nm. For the FND samples, individual spectra were acquired for ~30 FNDs and averaged. See SI Figure S4 for PL maps and examples of individual FND spectra. Three spectra were acquired in different locations of the bulk sample and also averaged. All samples show a pronounced NV$^0$ ZPL at 575 nm. The CVD bulk sample shows a clear NV$^-$ ZPL at 637 nm, while the NV$^-$ ZPL is much less pronounced in both FND samples. Qualitatively, the CVD FND sample shows a much stronger NV$^0$ PL signal compared to the CVD bulk sample, indicated by a higher PL signal in the 570-630 nm spectral region. The HPHT FND sample shows an NV$^0$ contribution between the CVD bulk and CVD FND sample.

To quantify these spectral differences, we employed a dual-wavelength-excitation protocol (DEP) recently reported by our team [27]. Briefly, PL spectra were acquired using 405 nm and 520 nm excitation to infer pure NV$^0$ and NV$^-$ spectra [30]. Pure NV$^-$ and NV$^0$ spectra were then used to determine the NV$^-$ ratio as the ratio between NV$^-$ PL and total (NV$^-$ and NV$^0$) NV PL using least-squares fitting for each spectrum. Figure 2b) shows a statistical analysis of the NV$^-$ ratio for all three samples. The CVD bulk sample exhibits a NV$^-$ ratio of 60% with minimal variation in the three different locations investigated. The CVD FNDs exhibit a significantly lower average NV$^-$ ratio of 31% with a large spread of values between 1 and 55%. In agreement with the qualitative observations in Figure 2a), the HPHT FNDs show an average NV$^-$ ratio of 46% with a slightly lower particle-to-particle variation between 25% and 64%.

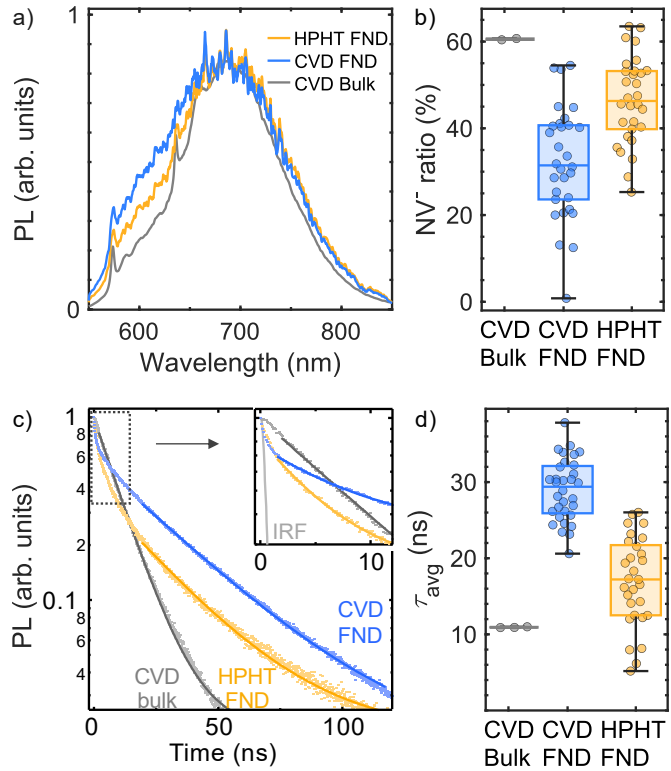

*Figure 2. Photoluminescence (PL) spectra and PL lifetime of CVD bulk, CVD FND and HPHT FND samples. a) Normalized average PL spectra acquired with a confocal PL microscope using 520 nm excitation (150 µW). For the FND samples, spectra were acquired for ~30 FNDs and averaged. Three spectra were acquired in different locations of the bulk sample and also averaged. b) Statistical analysis of the NV$^-$ ratio. The box plot shows the median NV$^-$ ratio, the upper and lower quartiles (shaded box area), and the maximum and minimum values (whiskers). Dots show individual measurements. c) Average time-resolved PL decay traces for all samples. Markers show experimental data, and solid lines are single exponential fits to the data. The inset shows a zoom into the first 12 ns of the decays and the instrument response function (IRF).*

To further elucidate the observed differences in the NV$^-$ ratio, we investigated the PL lifetime of the same particles and locations on the bulk sample analysed in Figures 2a) and 2b). Figure 2c) shows averaged time-resolved PL decay traces, where dots represent experimental data and solid lines single exponential fits to the data. The inset shows a zoom into the first 12 ns of the decay and the instrument response function (IRF). In Figure 2c, we investigate PL decay traces averaged over ~30 FNDs for each FND sample to identify systematic differences. Overall, the CVD bulk sample shows a faster decay than both FND samples. While the CVD bulk sample exhibits a predominantly single exponential decay with a lifetime of ~11 ns, both FND samples show a more complex multi-exponential decay. The latter is partially a result of the averaging over many individual particles. However, even individual particles generally exhibit a multi-exponential decay, which is much more pronounced in HPHT FNDs than in CVD FNDs (see SI Figure S5 for examples). Between 20 and 120 ns, both FND samples show a slow decay component with a lifetime of 34 ns (CVD FNDs) and 28 ns (HPHT FNDs) determined via a single exponential tail fit (blue and yellow solid lines in Figure 2c, respectively). Between 0 and 12 ns, the FND samples show two distinct decay constants, a very fast decay component in the first 1.5 ns with a lifetime of 0.6 ns, followed by a slower decay component between 1.5 and 12 ns with a time constant of 8.5 ns (CVD FNDs) and 5.7 ns (HPHT FNDs, see inset in Figure 2c). The fast decay component is not present in the CVD bulk sample, and we infer that it is related to NVs located very close to the surface of particles. Since the time constant and amplitude of the decay in the first 1.5 ns are similar for both FND samples, we will focus on the decay from 1.5-120 ns.

To quantitatively compare the overall PL decay dynamics, we employ a double-exponential fit to analyse the PL decay traces and calculate an amplitude-weighted average PL lifetime ($\tau_{avg}$) for each FND PL decay trace. Figure 2d) shows a box plot of $\tau_{avg}$ for both FND samples and for the three investigated locations on the CVD bulk sample. The CVD bulk sample exhibits a typical NV lifetime of 11 ns. The CVD FNDs and HPHT FNDs have a mean $\tau_{avg}$ of 29 ns and 17 ns with an interquartile range of 6.2 and 9.2, respectively.

Overall, the differences between the investigated samples shown in Figure 2 can be explained by considering (1) the electron donor and acceptor concentrations and (2) the sample form factor. Substitutional nitrogen is the main electron donor in all samples, and vacancy-related defects are the main electron acceptors. The CVD bulk sample has a $N_s^0$ and NV concentrations of 2.05 ppm and 0.27 ppm, respectively, and shows a NV$^-$ ratio of 60%. Milling causes a drastic increase in surface-to-volume ratio in the CVD FND particles. Electrons are trapped at surface-related electron acceptors, and fewer electron donors are available to create NV$^-$, resulting in the observed decrease in the NV$^-$ ratio in the CVD FND sample (Figures 2a and 2b). Due to the relatively low absolute concentration of $N_s^0$ in the CVD samples, the NV$^-$ ratio also becomes very sensitive to the FND particles' size and variations in surface chemistry – both of which affect the number of surface-related electron traps – and leads to a great variation in individual FND NV$^-$ ratios (Figure 2b) [31]. For instance, a 100 nm spherical particle with a typical surface defect density of order $10^{13}$ cm$^{-2}$ [32, 33] containing 2 ppm $N_s^0$, which corresponds to on the order of 1000 surface acceptors and 100 electron donors per particle, respectively, may not be expected to sustain a strong NV$^-$ population. The HPHT FND sample has much higher average $N_s^0$ and NV concentrations of over 100 ppm and 1 ppm, respectively, and hence around 100 electron donors per NV compared to only 7 donors per NV for the CVD samples. In moderately irradiated ($1\times10^{18}$ e$^-$ cm$^{-2}$) bulk HPHT diamonds, this large excess of electron donors leads to NV ratios of [34]. In the HPHT NDs investigated here, the higher irradiation dose ($\sim 1\times10^{19}$ e$^-$ cm$^{-2}$) increases the concentration of electron-accepting multi-vacancy defect [35] and the nanoscale particle size the number of surface-related electron traps. These opposing effects of higher electron donor and higher electron acceptor concentrations result in an average NV$^-$ ratio of the HPHT FNDs between CVD bulk and CVD FND samples.

In bulk diamond, the PL lifetime of NV$^-$ and NV$^0$ are 10-13 ns [36] and ~19 ns [37], respectively. In nanodiamonds, values increase significantly to above 20 ns (NV$^-$) and close to 30 ns (NV$^0$), respectively [38] due to a modulation in the local density of states by the dielectric properties of the particles' nanoscale environment [39]. Hence, the average lifetimes determined here for the bulk (11 ns) and FND CVD samples (29 ns) are in agreement with previous reports. The shorter average lifetime of 17 ns on HPHT FNDs is likely associated with higher $N_s^0$ concentrations [40] and higher non-radiative decay rates caused by vacancy-related defects [41]. The CVD FNDs show a stronger contribution of a longer decay component (Figure 2c), which could also be related to the higher relative NV$^0$ contribution. However, we find that the NV$^-$ ratio and the average PL lifetime are not correlated for the CVD FND particles (see SI Figure S6), suggesting that the lower (relative to the HPHT particles) $N_s^0$ concentration and vacancy defect concentration are the main contributors.

We then performed ODMR measurements to investigate the NV spin properties. The samples were placed on a strip line on a printed circuit board for MW delivery, and pulsed MW ODMR spectra were acquired with a custom-built confocal PL microscope using 514 nm laser excitation (see the Experimental section for details). Figure 3 a) shows typical ODMR spectra for all three samples and Figures 3b) to 3d) show a statistical analysis of the ODMR contrast, ODMR zero field splitting parameters *D* and *2E*. Overall, we observe a similar average ODMR contrast of ~3% for all samples (Figure 3b). The CVD bulk sample only shows a small variation in contrast between individual measurements, while the standard deviation between individual FNDs is 1.3 and 1.0 for the CVD FNDs and HPHT FNDs, respectively. Among randomly selected particles, we also find that a much higher fraction of the CVD nanoparticles does not produce a measurable ODMR signal (referred to as Low SNR) compared to the HPHT particles (Figure 3 b) left). This result is consistent with the measurements in Figure 2b), which show that ~25% of CVD FNDs show an NV$^-$ ratio below 25%. This means 75% of collected photons originate from NV$^0$, which likely explains the lack of ODMR contrast of many CVD FNDs. Again, this is likely due to the lower concentration of $N_s^0$ electron donors in these particles, leading to a greater sensitivity to acceptor-like surface states.

The impact of the milling process may be more apparent in the resonant zero-field ODMR frequencies *D* and *E*, which are sensitive to crystal strain. We fit each ODMR spectrum with two Lorentzian functions centered at frequencies $f_-$ and $f_+$, and from these fits, we calculate the parameters $D = \frac{f_+ + f_-}{2}$ and $2E = f_+ - f_-$. These parameters are sensitive to axial and transverse strain relative to the NV symmetry axis, respectively. Starting with $D$ (Figure 3c), we find a large spread in both FND data sets, with maximal deviations from typical bulk values of 2870 MHz of 2.5 MHz, translating to strain of approximately 200 MPa [42]. Laser-induced heating may also shift $D$ by 74k Hz K$^{-1}$ [43] but would lead to a similar shift for each FND. Milling, on the other hand, may create different amounts of strain in individual particles. Hence, we attribute the shift in $D$ in CVD and HPHT FND particles largely to milling-induced strain.

Figure 3d) shows the fit values of $2E$. Again, we observe a significant spread across the FND samples, with higher average values of 7.7 ± 1.9 (CVD FNDs) and 11.2 ± 1.6 (HPHT FNDs) compared to the bulk sample value of 5.1 ± 0.1. This time, however, the HPHT FNDs feature a significantly larger splitting than the CVD particles. As well as being sensitive to transverse strain [44], $E$ is also sensitive to charged defects within the diamond [45-47]. Hence, in this case, the difference between CVD and HPHT FNDs is likely due to the higher $N_s^0$ concentration and electron irradiation dose associated with the HPHT material.

Diamond crystal strain can also be assessed using birefringence measurements. We acquired birefringence maps of the bulk CVD diamonds, yielding average birefringence values of a difference in the refractive index between the two birefringent axes of $\Delta n$ = 4e$^{-5}$ to 9e$^{-5}$. The birefringence maps show locally resolved birefringence measurements, see SI Figure S7. The maps show a relatively homogeneous birefringence with only small regions with increased strain. While these regions might contribute some FNDs with increased strain, they don't explain the significant increase in spread and values of $2E$ in the CVD FNDs after milling. Therefore, we conclude that this increase is likely due to increased strain caused by the milling process itself.

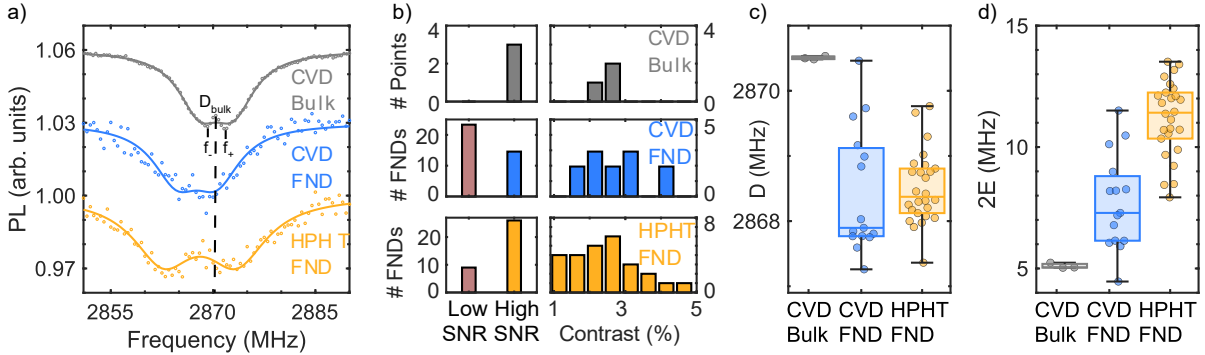

*Figure 3*. Analysis of the NV$^-$ optically detected magnetic resonance (ODMR) properties. a) Typical ODMR spectra for all samples. Dots show experimental data and solid lines double Lorentzian fits to the data. b) Histograms of ODMR signal-to-noise ratio (SNR, left) and ODMR contrast (right). c) and d) Box plot of ODMR zero field splitting parameters D (panel c) and 2E (panel d) determined from the double Lorenzian fits. Markers show individual measurements, the box center line the data mean, the shaded box area the interquartile range, and the whiskers the data minima and maxima.

For a final set of measurements, we consider $T_1$ relaxometry, which is becoming increasingly important for nanodiamond-based quantum sensing [48]. For these measurements, sensitivity is proportional to the base $T_1$ decay rate, which typically reduces from several ms in bulk samples to sub-100 µs values in NDs of varying sizes [49]. Several factors are known to impact NV $T_1$, including defect concentration [50-52] within the diamond and NV proximity to surface defects [53]. NDs sourced from a low-nitrogen parent crystal, then, may offer an advantage compared to HPHT particles.

Figure 4 shows $T_1$ decay traces collected from the three samples using a custom-built confocal PL microscope (see Experimental Section for details) and the laser excitation and MW pulse sequence (Figure 4a) used for the data acquisition. As shown in Figure 4a, all samples are excited with a 514 nm laser, and the traces are collected without MW ('Signal') and when MW is applied ('Reference'). 'Reference' is subtracted from 'Signal' to remove charge-state-related and other fluorescence artefacts [54]. To obtain reliable averaged results for the FND samples, we first acquired $T_1$ decay traces for particle ensembles, by selecting regions with a high density of FNDs within the layer (see SI Figure S4a,d). Figure 4b shows the resulting average $T_1$ decay traces for the FND samples compared to the decay of the CVD bulk sample. Both CVD bulk and CVD FND average show long $T_1$ decay times of 3.2 ± 0.7 and 4.7 ±1.6, respectively, determined by a single exponential fit to the data. The slightly shorter $T_1$ decay time observed for the bulk sample is likely an artefact caused by the collection of signal from regions of the bulk sample, where the NV spins throughout the collection volume may not be uniformly initialized. The HPHT FNDs show a much shorter $T_1$ decay time of 0.17 ± 0.01 ms. Hence, these measurements show that on average the milling process has a minimal impact on the $T_1$ relaxation times of NV centers inside the crystals. While most of the particles in the CVD FND sample have a diameter of ~ 100 nm (see Figure 1d), a broad range of particle sizes are present in this sample and it is possible that larger particles dominate the CVD FND decay trace shown in Figure 4b. Therefore, we also investigated a smaller individual particle with a diameter well below 250 nm (determined via the point-spread function of a diffraction-limited 514 nm laser beam) using a confocal PL microscope to determine if the long average $T_1$ decay times can be preserved in single FNDs. Figure 4c shows the $T_1$ decay for a single FND (red trace) compared to the CVD FND average and HPHT FND decays in the 0-1 ms delay time range. While the data is noisy due to the lower PL signal in the CVD FNDs compared to HPHT FNDs, a single exponential fit to the data clearly shows that the PL intensity decreases only by about 10% during the first millisecond of the decay, similar to the CVD FND average decay. During this time, the HPHT FND PL decreases to zero. This suggests that bulk-like $T_1$ decay times above 3 ms can be preserved in milled CVD FNDs with a tailored concentration of $N_s^0$ and NV centers.

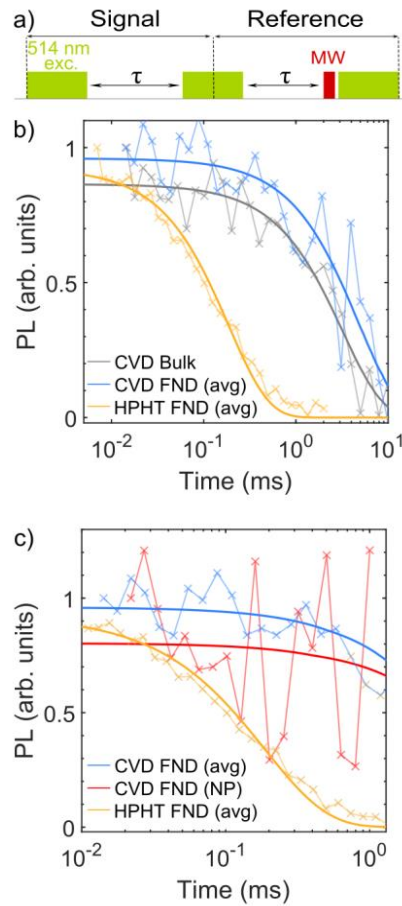

*Figure 4. Relaxation time $T_1$ for the three samples. a) Acquisition scheme in which a 514 nm laser is used as excitation source. The excitation is followed by a dark time τ and collection (signal), the next pulse and dark time are followed by microwave delivery and readout (reference). b) $T_1$ comparison for CVD bulk and FND samples between 0 and 10ms, inset shows the decay in the first ms. c) $T_1$ comparison of nanoparticles. Single particle CVD FND (NP) is compared against the average relaxation time of CVD FND (avg). Markers show experimental data, and solid lines are single exponential fits to the data.*

**Summary**
Our study demonstrates that ball milling of nitrogen-doped CVD single crystal diamond with intermediate $N_s^0$ and $NV^-$ concentrations on the order of ~2 ppm and ~0.2 ppm yields FNDs with bulk-like $T_1$ spin relaxation times and tuneable NV properties. We find that CVD FNDs show significantly lower $NV^-/NV^0$ ratios (31%) compared to bulk CVD (60%) and HPHT FNDs (46%), likely caused by the lower concentration of $N_s^0$ electron donors in the CVD samples compared to HPHT, resulting in a stronger influence of the dielectric environment on the NV charge state in CVD FNDs. Average PL lifetimes increase from 11 ns (bulk CVD) to 29 ns (CVD FNDs), likely caused by a combination of a lower effective refractive index leading to slower spontaneous emission and a higher relative contribution of $NV^0$ to the overall PL. The HPHT FNDs show a shorter average lifetime of 17 ns than the CVD FNDs, mainly caused by the higher concentration of other charged defects in the HPHT FNDs that lead to a shorter NV excited state lifetime.

We find that the ODMR contrast is comparable across all samples (~3%), while more than half of the CVD FNDs show no ODMR signal at all, likely due to the lower $NV^-$ content in these particles. Milling introduces strain, evident from the shift in the zero-field splitting D (~2.5 MHz deviation from bulk) and increased 2E parameter, which is smaller for CVD FNDs than HPHT FNDs, likely due to lower charged defect concentrations. The CVD FNDs retain long $T_1$ relaxation times (4.7 ± 1.6 ms), comparable to bulk CVD (3.2 ± 0.7 ms) and significantly longer than HPHT FNDs (0.17 ± 0.01 ms). Single nanoparticle measurements confirm that long $T_1$ times can also be preserved in sub-250 nm particles.

While challenges remain in stabilizing the $NV^-$ charge state in milled CVD nanodiamonds with $N_s^0$ concentrations of ~1 ppm, the results highlight a promising pathway toward scalable production of FNDs with quantum-grade spin coherence properties for sensing applications. Future work will focus on optimizing surface chemistry and defect engineering to increase $NV^-$ charge state stability without compromising coherence.

**Conflicts of interest**
There are no conflicts to declare.

**Data availability**
All data needed to evaluate the conclusions in the paper are present in the paper and/or the ESI. Any additional datasets, analysis details, and sample fabrication recipes are available upon request from the corresponding authors.

**Acknowledgments**
PR acknowledges support through an Australian Research Council (ARC) DECRA Fellowship (DE200100279) and an RMIT University Vice-Chancellor's Senior Research Fellowship. This work was supported by ARC Discovery Projects (DP220102518, DP250100125). This work was performed in part at the RMIT Microscopy and Microanalysis Facility (RMMF). AK acknowledges support by the DFG (FOR1493, project number 164674487) and the BMBF in the Cluster4Future QSENS. This material is based upon work supported by U.S. Department of Energy, Office of Science, Office of Fusion Energy Sciences and Office of Basic Energy Sciences under Award No. LAB 21-2491

Alessandro Mameli[1], Giannis Thalassinos[1], Marco Capelli[1], Johannes Ackermann[2], Edwin Mayes[1], Hiroshi Abe[3], Takeshi Ohshima[3,4], Tingpeng Luo[4], Volker Cimalla[5], Peter Knittel[5], Brant Gibson[1], Jan Jeske[5], Nikolai Dontschuk[6], Anke Krueger[2,7], Alastair Stacey,[1,8] Alexander Healey[1]*, Philipp Reineck[1]*

[1] School of Science, RMIT University, Melbourne, VIC 3001, Australia
[2] Institute of Organic Chemistry, University of Stuttgart, Pfaffenwaldring 55, 70569, Stuttgart, Germany
[3] National Institutes for Quantum and Radiological Science and Technology, Takasaki, Gunma, 370-1292, Japan
[4] Department of Materials Science, Tohoku University, Aoba, Sendai, Miyagi 980-8579, Japan
[5] Fraunhofer Institute for Applied Solid State Physics IAF, Tullastraße 72, 79108 Freiburg im Breisgau, Germany
[6] School of Physics, University of Melbourne, Parkville, Victoria 3010, Australia
[7] Center for Integrated Quantum Science and Technology (IQST), 70569, Stuttgart, Germany
[8] Princeton Plasma Physics Laboratory, Princeton University, Princeton, NJ 08540, USA

* *alexander.healey2@rmit.edu.au*, philipp.reineck@rmit.edu.au

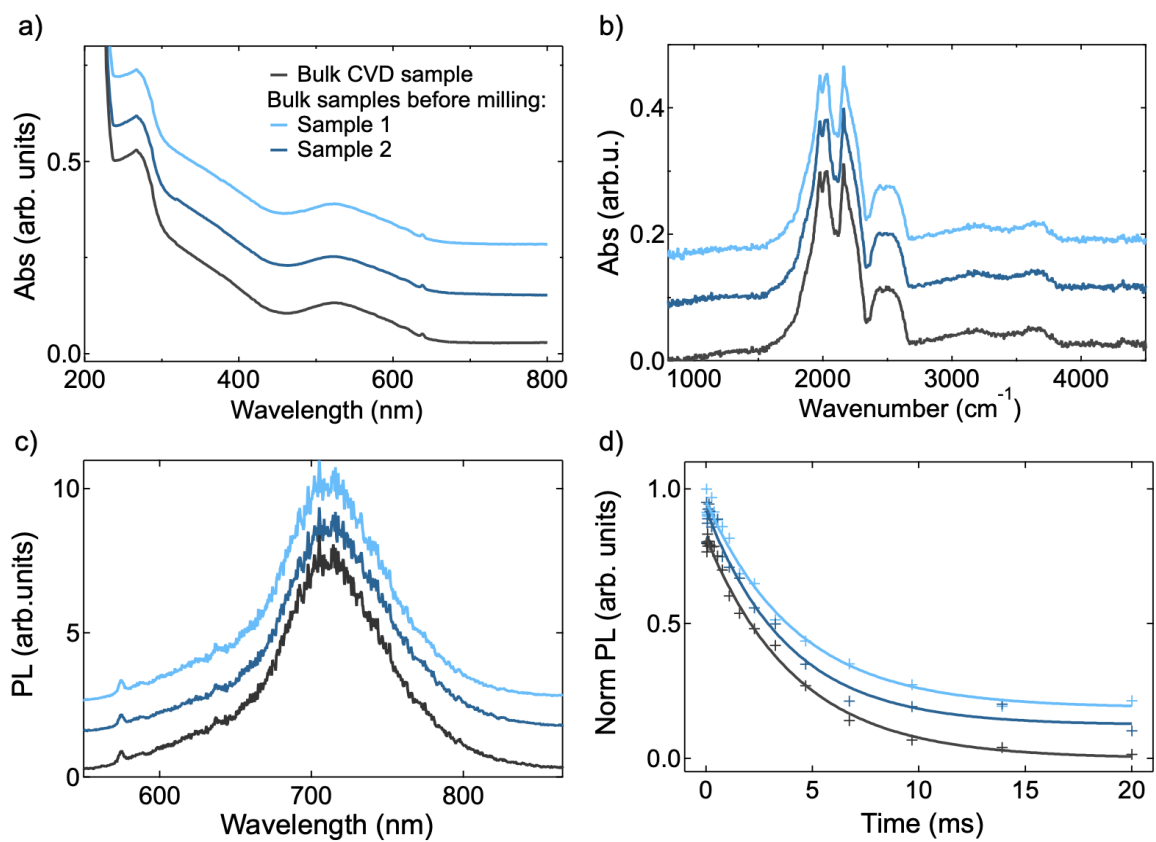

**Figure S1**. Materials and fluorescence properties of all three CVD bulk diamond samples after irradiation and annealing. The spectra are offset by 10% for clarity. a) UV-Vis absorption spectra. b) FTIR absorption spectra. c) PL spectra. d) NV T1 spin relaxation traces.

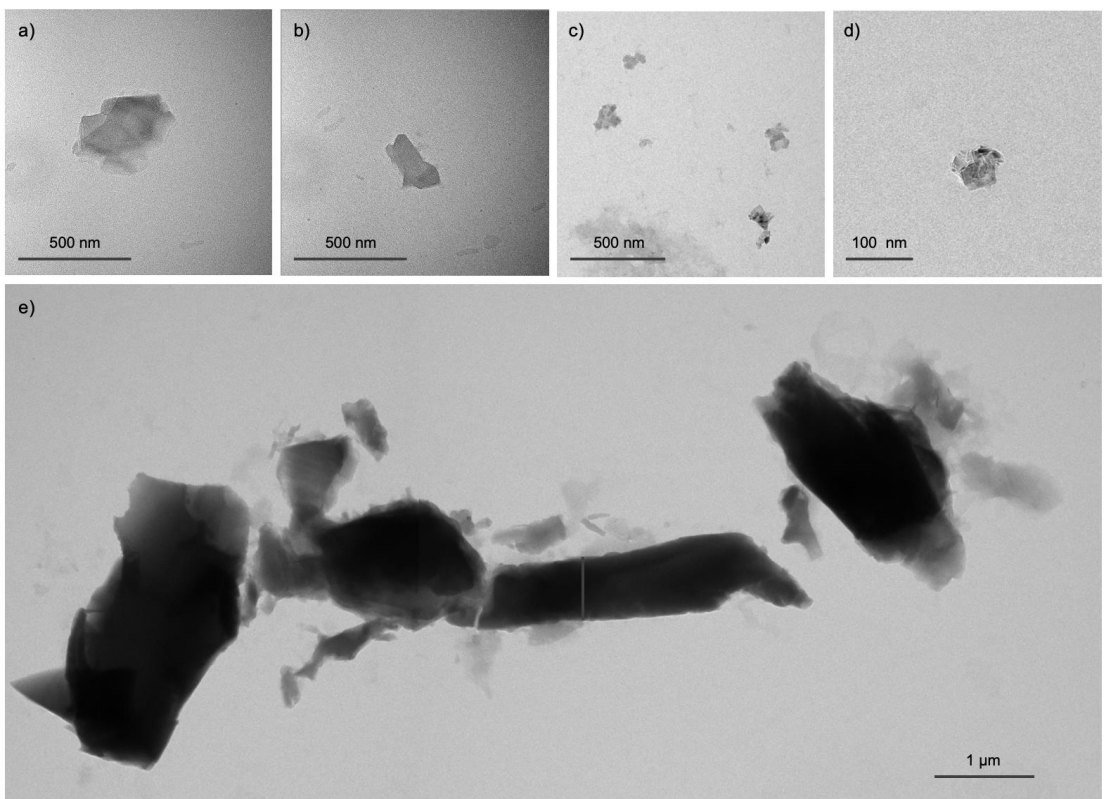

**Figure S2**. TEM images of CVD FND particles at different magnifications. The top row (panels a-d) shows images of typical FNDs based on the dynamic light scattering particle size analysis. e) Image of several larger micron-sized diamond particles surrounded by smaller particles.

**Custom-built confocal system**

The system is designed to operate with various excitation sources, including a tuneable laser (Fianium WhiteLase 400-SC, NKT Photonics) operating within the 400–630 nm wavelength range, typically set at 520 nm with a repetition rate of 5–80 MHz. Additional excitation sources include a 405 nm diode laser (Coherent Obis FP 1193825, Edmund Optics) and a continuous-wave (CW) 514 nm laser (Cobolt 06-MLD, Hubner Photonics). Depending on the selected excitation source, an appropriate set of excitation and collection optical filters is employed.

The excitation beam is fiber-coupled to a collimator and directed through a neutral density wheel (NDC-50C-4M, Thorlabs) to modulate the excitation power. The power level is measured using a power meter (PM100D, Thorlabs) along the beam path. The collimated beam then passes through a 532 nm short-pass filter (BSP01-532R-25, Semrock), ensuring transmission of only the excitation wavelength.

The beam then proceeds to a 532 nm dichroic beam-splitter (LPD02-532RU-25, Semrock) and is directed toward a piezoelectric scanning stage (P-545.xC8S, Physik Instrumente), with a travel range of 200 μm along each axis. The beam is then focused onto the sample using a 100× objective lens (NA = 0.9), mounted on the scanning stage. Fluorescence emitted from the sample is collected by the same objective and reflected by the dichroic beam splitter toward the detection path.

A 532 nm long-pass filter (BLP01-532R-25, Semrock) is used to block residual excitation light, allowing only the fluorescence signal originating from the diamond sample to be transmitted. The fluorescence is then coupled into a 62.5 μm core diameter multimode optical fiber serving as a pinhole. The collected signal is split using a 75:25 fiber optic beam splitter: 25% of the signal is directed to an avalanche photodiode (APD; SPCM-AQRH-14-TR, Excelitas) for fluorescence imaging, while the remaining 75% is sent to a spectrometer (SpectraPro SP-2500, Princeton Instruments) coupled to a charge-coupled device (CCD) camera (PIXIS 100BR, Princeton Instruments) for fluorescence spectral acquisition.

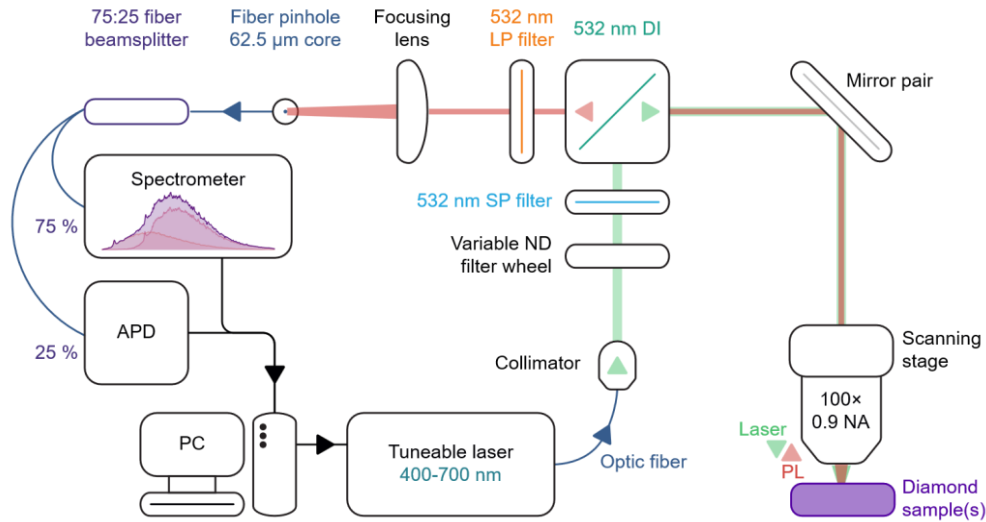

**Figure S3**. Schematic illustration of the custom-built confocal PL setup used for the characterization of the NV PL and spin properties. A laser source was fiber-coupled to a collimator, which directed the beam through a series of optical filters, including short-pass and neutral density filters, before reaching a dichroic mirror. The dichroic mirror reflected the excitation beam into an air objective (NA = 0.9) mounted on a XYZ scanning stage. The resulting PL emission was collected by the same objective and transmitted back through the dichroic mirror to a multimode optical fiber. The collected signal was then split by a beam splitter, with 75% directed to a spectrometer and 25% to an avalanche photodiode (APD) for detection. For time-resolved spectroscopy, a correlator card was used.

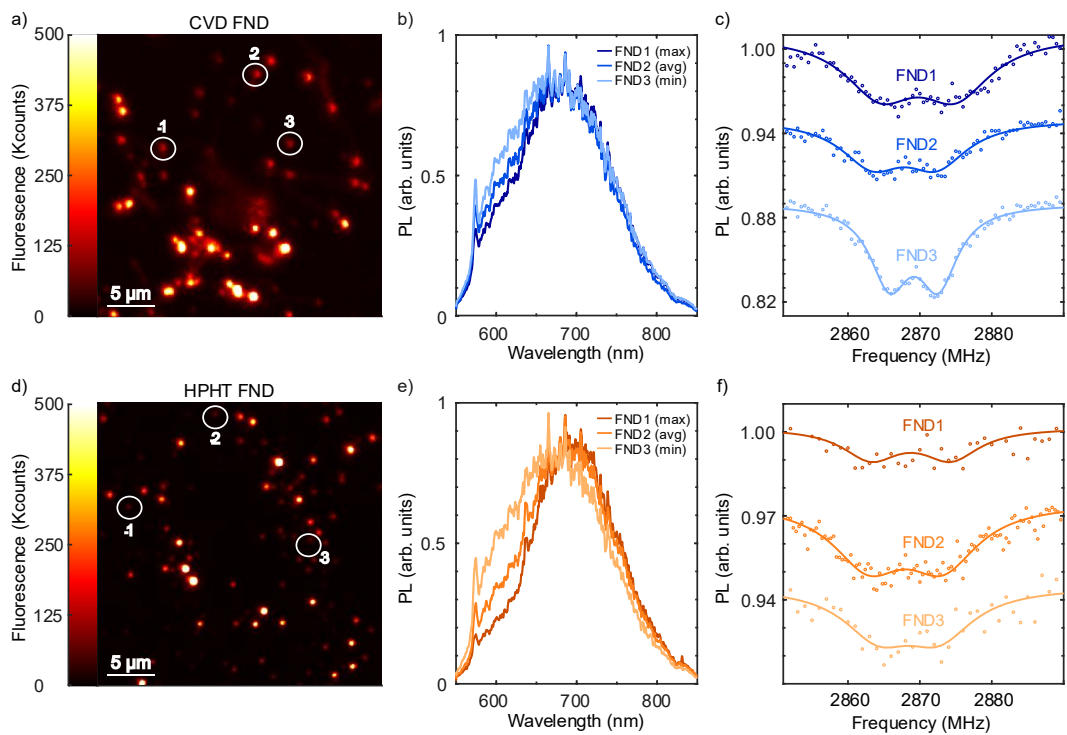

**Figure S4**. Analysis of three single particles of CVD FND (a-c) and HPHT FND (d-f). a) and d) PL map of FND acquired with a confocal PL microscope using 520 nm excitation (150 µW) and 532 nm SP/LP filtering. b) and e) Normalized PL spectra of particles selected according to their maximum (FND1), average (FND2) and minimum (FND3) NV$^-$ content. c) and f) ODMR spectra for selected FNDs. Measurements were acquired using a confocal PL microscope using 514 nm excitation (150 µW) and 532 nm SP and 633 nm LP filtering. Dots show experimental data and solid lines double Lorentzian fits to the data. The spectra are offset by 3% for clarity.

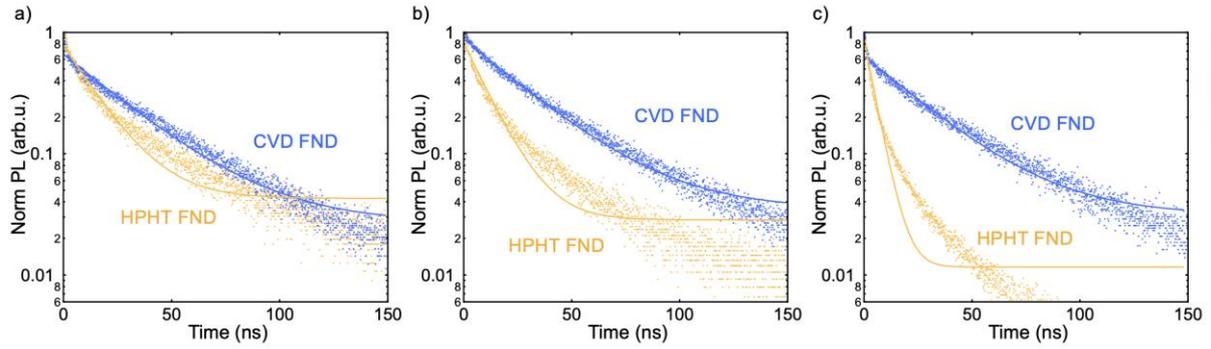

**Figure S5**. Time-resolved PL decay traces for three different HPHT and CVD FND particles. Dots are experimental data and solid lines are single exponential fits to the data. The three particles shown here are not the same particles shown in the previous figure. While single exponential decays do not fit the decays of CVD and HPHT FNDs, the HPHT FNDs more clearly show a multi-exponential decay than the CVD FNDs.

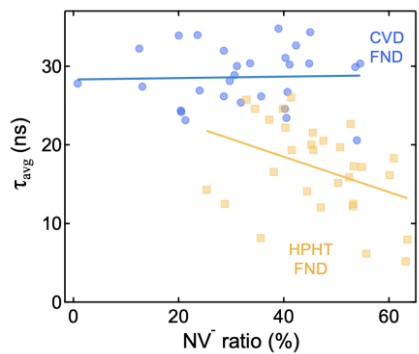

**Figure S6**. Scatter plot of the average lifetime vs. NV⁻ ratio for individual FND measurements. The solid traces are linear fits to the experimental data.

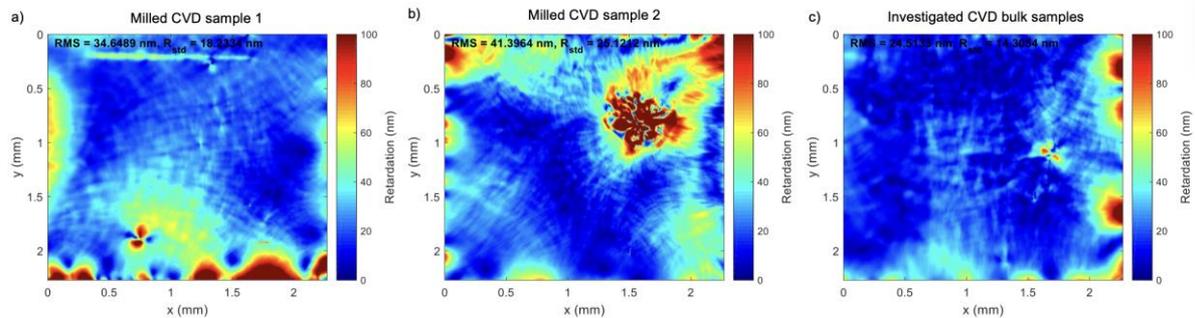

**Figure S7**. Birefringence map of the two CVD samples milled to nanodiamonds (a, b) and the CVD bulk samples investigated in the main text.

**ODMR data acquisition and analysis**

We performed ODMR measurements using a CW protocol for simplicity, normalizing against a no-MW measurement at each frequency and not necessarily optimizing relative laser and MW powers for maximal readout contrast. $T_1$ measurements were performed with relatively high laser power (150 μW) and long laser pulse durations (exceeding 10 μs) to ensure good NV spin initialization. For the low-nitrogen samples, where NV charge state was observed to be dependent on laser power and duty cycle, the recorded PL (taken from a 1 us window at the start of the laser pulse) was normalized against a "reference" window at the back of the pulse. This procedure is partially responsible for the higher levels of relative noise apparent in the associated measurements shown in Fig. 4b,c. For all samples (as depicted in Fig. 4a), the data presented is the difference between a sequence including only dark time between laser pulses and one with a MW π-pulse just prior to readout. This procedure normalizes against dark ionization processes and laser power fluctuations.

ODMR data analysis was performed using MATLAB R2024. Experimental data were fitted using a bi-Lorentzian model described by the following equation:

$$y(f) = 1 - \frac{A_-}{1 + \left(\frac{f - f_-}{\Gamma_+}\right)^2} - \frac{A_+}{1 + \left(\frac{f - f_+}{\Gamma_+}\right)^2}$$

in where $A_-, A_+$ are the amplitudes, $f_-, f_+$ are the center frequencies, and $\Gamma_-, \Gamma_+$ are full-width at half-maximum (FWHM) of each dip.

To constrain the fitting into a symmetrical curve, boundary conditions were applied such that $A_- = A_+$ and $\Gamma_- = \Gamma_+$ for both dips of the ODMR curve. For each of the samples, the D and 2E parameters were calculated according to the equations $D = \frac{f_+ + f_-}{2}$ and $2E = f_+ - f_-$. The statistical error for the computed parameters was calculated through the covariance matrix derived from the Jacobian of the fit.